\documentclass[a4paper,11pt]{article}
\pdfoutput=1 

\usepackage{jinstpub} 

\usepackage{caption}
\usepackage{subcaption}
\usepackage{verbatim}
\usepackage[symbol]{footmisc}
\usepackage[normalem]{ulem}
\newcommand\redout{\bgroup\markoverwith
{\textcolor{red}{\rule[0.5ex]{2pt}{0.8pt}}}\ULon}

\title{\boldmath Performance of PMTs for the JSNS$^{2}$ experiment}

\author[1]{\normalsize{J.~S.~Park}}
\author[1]{H.~Furuta}
\author[1]{T.~Maruyama}
\author[1]{S.~Monjushiro}
\author[1]{K.~Nishikawa\footnote[2]{deceased}} 
\author[1]{M.~Taira}

\author[2]{J.~S.~Jang\footnote[1]{Corresponding author : jsjang@gist.ac.kr}}

\author[3]{K.~K.~Joo}
\author[3]{J.~Y.~Kim}
\author[3]{I.~T.~Lim}
\author[3]{D.~H.~Moon}
\author[3]{J.~H.~Seo}
\author[3]{C.~D.~Shin}
\author[3]{A.~Zohaib}
\author[3]{P.~Gwak}
\author[3]{M.~Jang}

\author[4]{S.~Ajimura} 
\author[4]{T.~Hiraiwa} 
\author[4]{T.~Nakano} 
\author[4]{M.~Nomachi} 
\author[4]{T.~Shima} 
\author[4]{Y.~Sugaya}

\author[5]{M.~K.~Cheoun}

\author[6]{J.~H.~Choi}
\author[6]{M.~Y.~Pac}

\author[7]{T.~Dodo}
\author[7, 8]{Y.~Hino}
\author[7]{F.~Suekane}
\author[7]{R.~Ujiie}

\author[8]{M.~Harada} 
\author[8]{S.~Hasegawa} 
\author[8]{Y.~Kasugai} 
\author[8]{S.~Meigo} 
\author[8]{K.~Sakai} 
\author[8]{S.~Sakamoto}
\author[8]{K.~Suzuya}

\author[9]{J.~R.~Jordan} 
\author[9]{J.~Spitz} 
\author[9]{E.~Marzec} 
\author[9]{M.~Botran}

\author[10]{T.~Kawasaki} 
\author[10]{T.~Konno}

\author[11]{H.~I.~Jang}

\author[12]{S.~K.~Kang}

\author[13]{E.~J.~Kim}

\author[14]{H.~Seo} 
\author[14]{S.~Y.~Kim}

\author[15]{W.~Kim}

\author[16]{M.~Niiyama}

\author[17]{S.~J.~M.~Peeters}

\author[18]{H.~Ray}

\author[19]{C.~Rott} 
\author[19]{I.~Yu} 
\author[19]{H.~Jeon}
\author[19]{S.~Jeon}
\author[19]{D.~E.~Jung} 
\author[19]{S.~B.~Kim}
\author[19]{E.~Kwon}
\author[19]{D.~H.~Lee}

\author[20]{I.~Stancu}

\author[21]{M.~Yeh}

\affiliation[1]{\it{\footnotesize{High Energy Accelerator Research Organization (KEK), Tsukuba, Ibaraki, JAPAN}}}
\affiliation[2]{\it{\footnotesize {GIST College, Gwangju Institute of Science and Technology, Gwangju, 61005, KOREA}}}
\affiliation[3]{\it{\footnotesize {Department of Physics, Chonnam National University, Gwangju, 61186, KOREA}}}
\affiliation[4]{\it{\it{\footnotesize {Research Center for Nuclear Physics, Osaka University, Osaka, JAPAN}}}}
\affiliation[5]{\it{\footnotesize {Department of Physics, Soongsil University, Seoul 06978, KOREA}}}
\affiliation[6]{\it{\footnotesize{Department of Radiology, Dongshin University, Chonnam 58245, KOREA}}}
\affiliation[7]{\it{\footnotesize {Research Center for Neutrino Science, Tohoku University, Sendai, Miyagi, JAPAN}}}
\affiliation[8]{\it{\footnotesize {J-PARC Center, JAEA, Tokai, Ibaraki JAPAN}}}
\affiliation[9]{\it{\footnotesize {University of Michigan, Ann Arbor, MI, 48109, USA}}}
\affiliation[10]{\it{\footnotesize{Department of Physics, Kitasato University, Sagamihara 252-0373, Kanagawa, JAPAN}}}
\affiliation[11]{\it{\footnotesize {Department of Fire Safety, Seoyeong University, Gwangju 61268, KOREA}}}
\affiliation[12]{\it{\footnotesize{School of Liberal Arts, Seoul National University of Science and Technology, Seoul, 139-743, KOREA}}}
\affiliation[13]{\it{\footnotesize {Division of Science Education, Physics major, Chonbuk National University, Jeonju, 54896, KOREA}}}
\affiliation[14]{\it{\footnotesize{Department of Physics and Astronomy, Seoul National University, Seoul 08826, KOREA}}}
\affiliation[15]{\it{\footnotesize{Department of Physics, Kyungpook National University, Daegu 41566, KOREA}}}
\affiliation[16]{\it{\footnotesize{Department of Physics, Kyoto Sangyo University, Kyoto, JAPAN}}}
\affiliation[17]{\it{\footnotesize{Department of Physics and Astronomy, University of Sussex, Brighton,UK}}}
\affiliation[18]{\it{\footnotesize {University of Florida, Gainesville, FL, 32611, USA}}}
\affiliation[19]{\it{\footnotesize{Department of Physics, Sungkyunkwan University, Suwon 16419, KOREA}}}
\affiliation[20]{\it{\footnotesize {University of Alabama, Tuscaloosa, AL, 35487, USA}}}
\affiliation[21]{\it{\footnotesize {Brookhaven National Laboratory, Upton, NY, 11973-5000, USA}}}  

\abstract{The JSNS$^{2}$ (J-PARC Sterile Neutrino Search at J-PARC Spallation Neutron Source) experiment aims to search for neutrino oscillations over a 24\,m short baseline at J-PARC. The JSNS$^{2}$ inner detector is filled with 17 tons of gadolinium-loaded liquid scintillator (LS) and both the intermediate $\gamma$-catcher and the optically separated outer veto are filled with un-loaded LS. Optical photons from scintillation are observed by 120 Photomultiplier Tubes (PMTs). A total of 130 PMTs for the JSNS$^{2}$ experiment were both donated by other experiments and purchased from Hamamatsu. Donated PMTs were purchased around 10 years ago, therefore JSNS$^{2}$ did pre-calibration of the PMTs including the purchased PMTs. 123 PMTs demonstrated acceptable performance for the JSNS$^{2}$ experiment, and 120 PMTs were installed in the detector.}


\begin{document}
\maketitle
\flushbottom

\section{Introduction}
The JSNS$^{2}$ (J-PARC Sterile Neutrino Search at J-PARC Spallation Neutron Source) experiment will search for neutrino oscillations at short baseline with $\Delta$m$^{2}$ near 1\,eV$^{2}$~\cite{cite:JSNS2_proposal}. JSNS$^{2}$ is located in the J-PARC Material and Life Science Experimental Facility (MLF). A 3\,GeV proton beam is delivered to a mercury target in the MLF, producing an intense source of muon antineutrinos from $\mu$+ decay-at-rest. JSNS$^{2}$ will search for $\overline{\nu}_{\mu}$ $\rightarrow$ $\overline{\nu}_{e}$ oscillations, which can be detected via inverse beta decay, $\overline{\nu}_{e} + p \rightarrow e^{+} + n$. The detector is comprised of an inner target volume, an intermediate $\gamma$-catcher region, and an optically separated outer veto~\cite{cite:JSNS2_TDR, cite:JSNS2_Veto_detector}. The target volume is filled with 17 tons of gadolinium-loaded liquid scintillator (LS) donated by the Daya Bay experiment~\cite{cite:DayaBay_GdLS}. Both the $\gamma$-catcher and veto are filled with 31 tons of LS. An event in the detector generates optical photons via scintillation and those photons are observed by 120 R7081 Hamamatsu 10-inch Photomultiplier Tubes (PMTs). Basic specifications for the Hamamatsu R7081 PMT are available in Ref~\cite{cite:R7081} and are summarized in Table.~\ref{tab:PMT}. 67 PMTs were donated by the RENO experiment and 24 PMTs were donated by the Double Chooz experiment and by the JSNS$^{2}$ collaborators. 39 PMTs were newly purchased from the Hamamatsu Photonics. JSNS$^{2}$ performed PMT pre-calibration to confirm the quality of the donated PMTs as well as the purchased PMTs. The pre-calibration results are compared with a data sheet that is provided by Hamamatsu Photonics. All but a few PMTs demonstrated acceptable performance for the JSNS$^{2}$ experiment and 120 PMTs were selected for installation in the detector.  

\begin{table}
\begin{center}
\begin{tabular}{c | c | c | c | c}
\hline
 Gain & Applied voltage & Quantum efficiency& Dark count & Peak to valley \\ 
         & for typical gain  & at 390\,nm & After 15 hours & ratio \\ \hline 
 Typ.       & Typ. (V) & Typ. (\%) & Typ. (s$^{-1}$), Max. (s$^{-1}$)& Min, Typ. \\ \hline
  1.0 $\times$ 10$^{7}$ & 1500 & 25 & 7000, 15000 & 1.5, 2.8 \\ \hline        
\end{tabular}
\end{center}
\caption{\setlength{\baselineskip}{4mm}Basic specifications of the R7081 10-inch PMT from the Hamamatsu data sheets.}
\label{tab:PMT}
\end{table}

\section{Pre-calibration setup}
\begin{figure}[h]
\begin{center}
\includegraphics[scale=0.5]{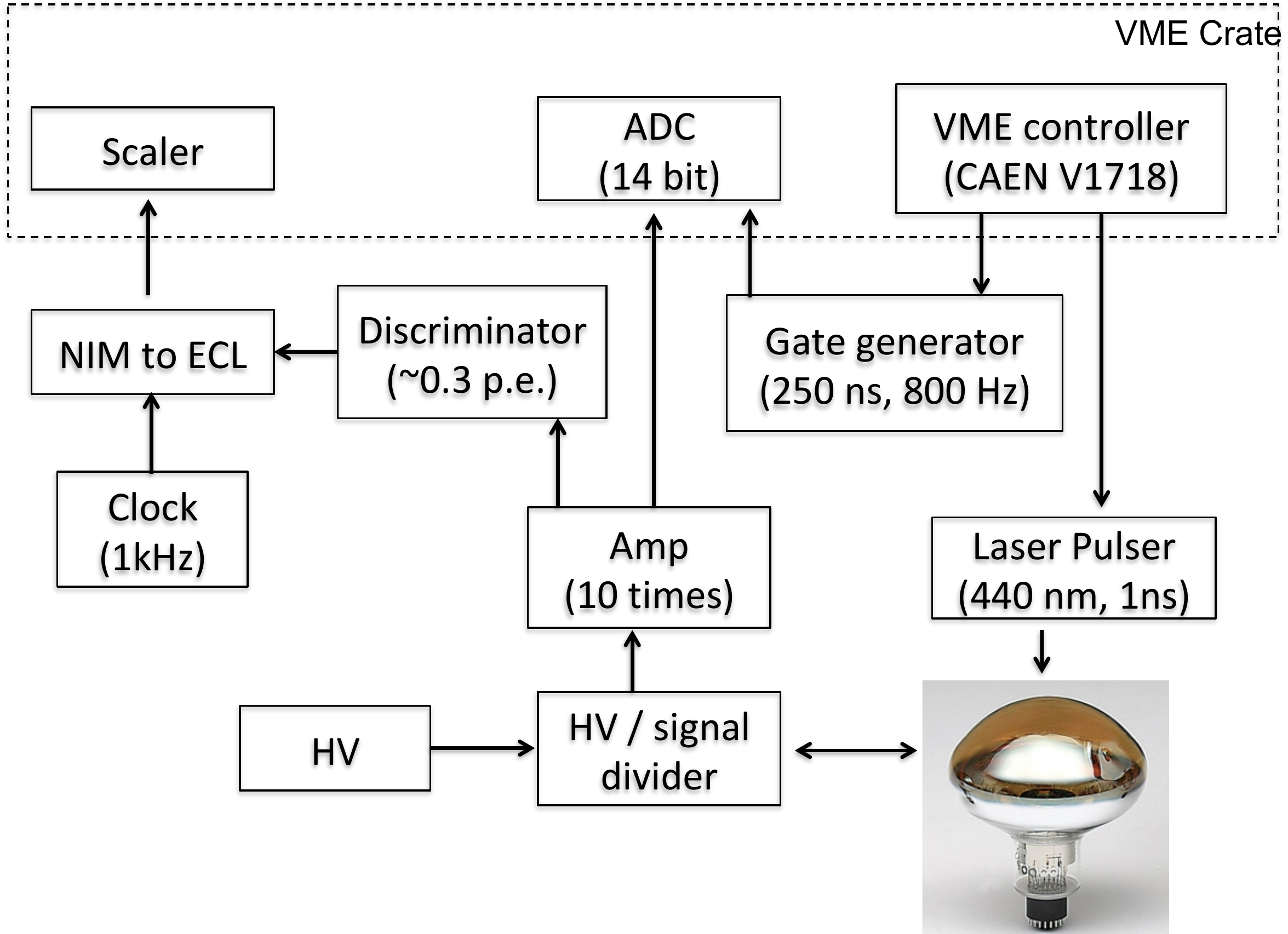}
\end{center}
\caption{\setlength{\baselineskip}{4mm}Schematic drawing of the data acquisition setup. The dashed block indicates a VME crate. The VME controller generates triggers at 800\,Hz for the laser pulser and the gate generator. A splitter supplies HV to the PMT and extracts the analog waveform. The analog signal is amplified by a factor of ten and inserted into a 14-bit ADC and a discriminator. A scaler measures the dark count rate and an ADC digitizes the analog signal to measure the PMT gain.}
\label{fig:setup}
\end{figure}

\begin{figure}[h]
\begin{center}
\includegraphics[scale=0.7]{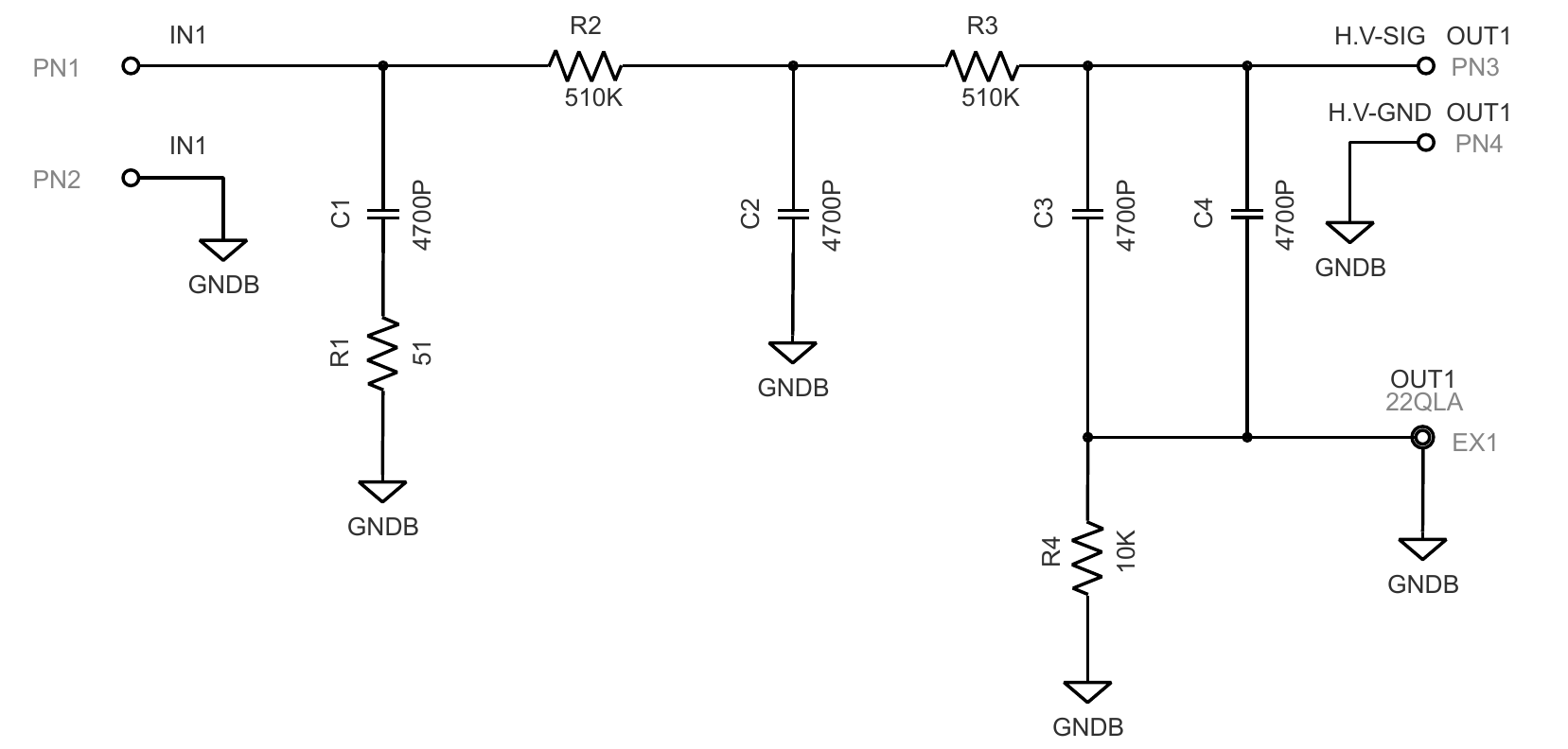}
\end{center}
\caption{\setlength{\baselineskip}{4mm}Circuit diagram of the splitter.}
\label{fig:splitter_circuit}
\end{figure}

For our pre-calibration the dark box originally used for the Double Chooz experiment's pre-calibration~\cite{cite:DC_PMT} was reused. The dark box is designed to hold up to 8 PMTs at once. Each PMT is mounted vertically and surrounded with a $\mu$-metal shield to remove any possible bias from external magnetic fields. A laser module which produces 1\,ns wide pulses of 440\,nm wavelength light is used to stimulate the PMTs. A VME controller is used to set the pulse rate to 800\,Hz, and a 1-to-8 laser splitter is used to distribute the single laser output to all 8 PMTs in the dark box. The laser light is injected into the center of the PMT glass using optical fibers. The analog output of each PMT is amplified by a factor of ten and split into two outputs. One output goes to a 14-bit analog-to-digital converter (ADC) with a 250\,ns window used to calculate the PMT charge, and the other goes to a counter module after passing through a discriminator to measure the dark count rate. The discriminator threshold is set to around 0.3\,p.e. equivalent at a gain of 10$^{7}$. Figure~\ref{fig:setup} shows the schematic flow of the data acquisition (DAQ) system. A single cable attached to a splitter circuit is used to supply high voltage (HV) to each PMT and to extract the PMT analog signal. Figure~\ref{fig:splitter_circuit} shows the circuit diagram of the splitter.

\section{Pre-calibration procedure}
\subsection{Eye inspection and noise check}
\begin{figure}[h]
\begin{center}
\includegraphics[scale=0.7]{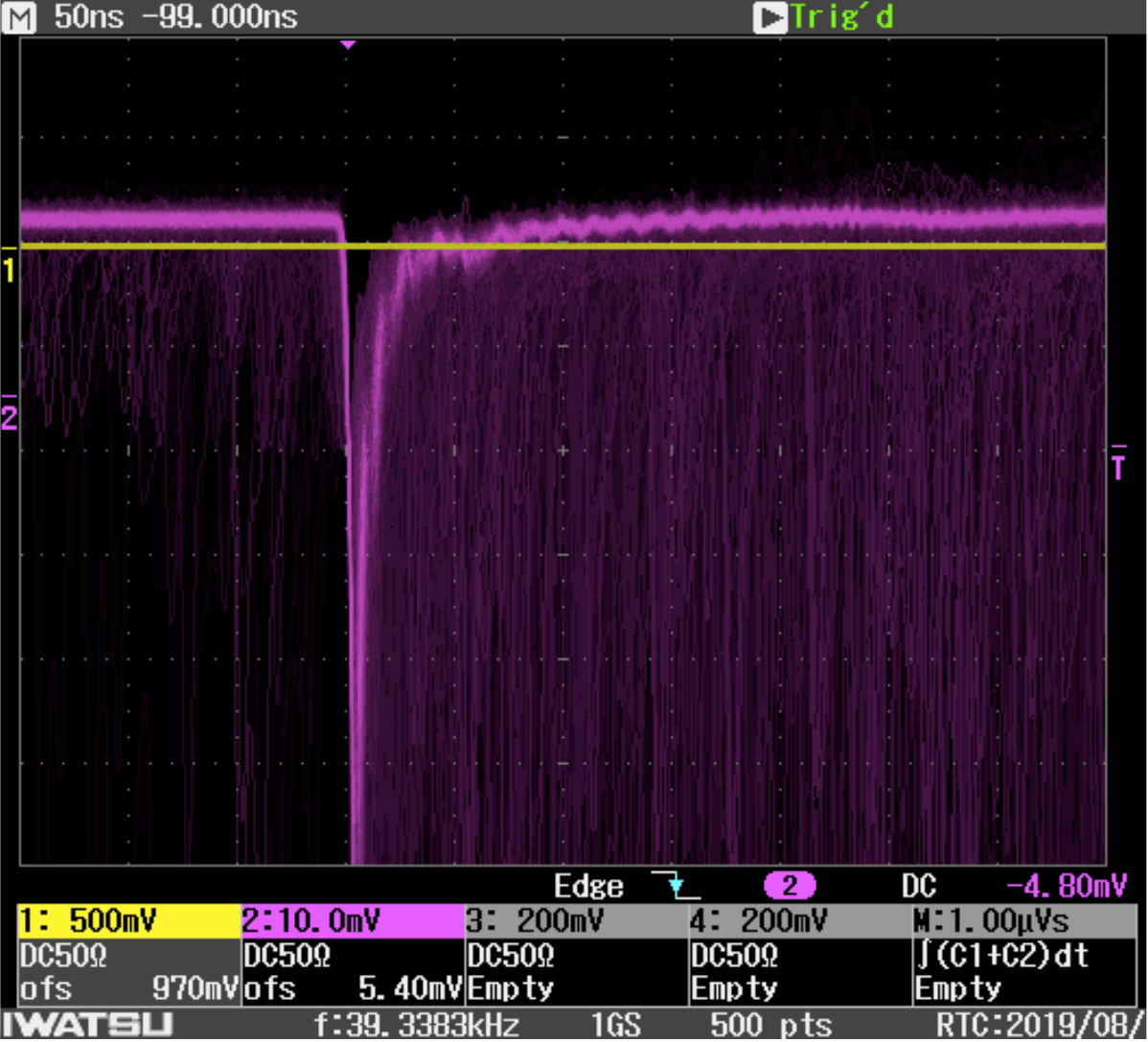}
\end{center}
\caption{\setlength{\baselineskip}{4mm}An accumulated oscilloscope screenshot of the noisy PMT. There are clear noisy signals. Note that the yellow trace is not relevant.}
\label{fig:noisy_PMT}
\end{figure}

The first step in the pre-calibration procedure is to identify malfunctioning PMTs. To do this, we applied 1400\,V to each PMT and checked the PMT analog signal with an oscilloscope. One PMT was found to have unacceptable noise level due to flashes of light produced by the PMT's base and was discarded. Figure~\ref{fig:noisy_PMT} shows an oscilloscope screenshot of the noisy PMT. After checking for noisy PMTs, we waited 1 hour to stabilize the PMT performance before taking gain measurement data. Note that 6 PMTs were excluded due to the different type of photocathode and operating method after the eye inspection.

\subsection{Gain curve}
\begin{figure}[h]
\begin{tabular}{c c}

\begin{subfigure}{.5\linewidth}
\centering
\includegraphics[scale=0.3]{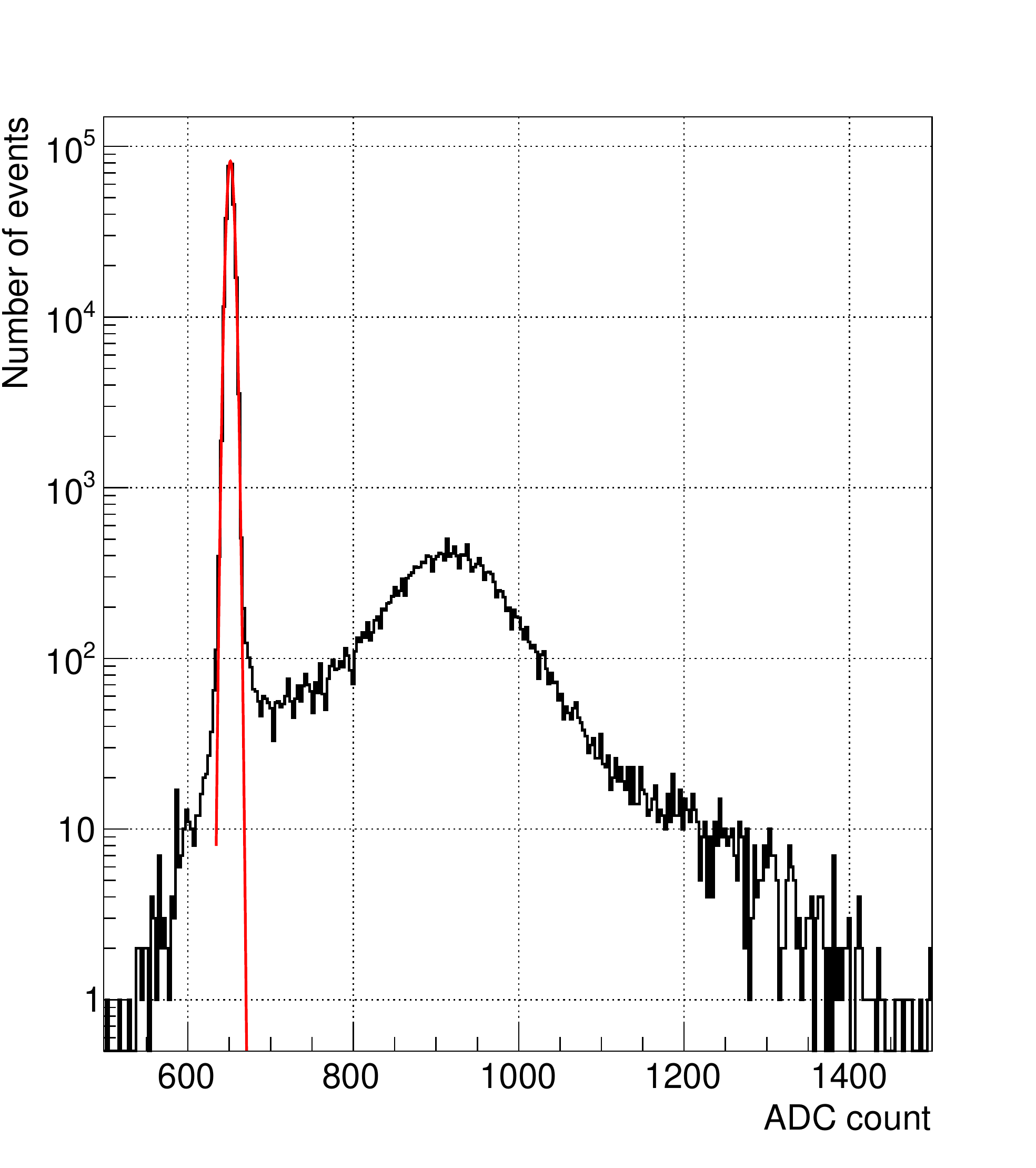}
\caption{With a pedestal fit.}
\label{fig:spe_a}
\end{subfigure}
&
\begin{subfigure}{.5\linewidth}
\centering
\includegraphics[scale=0.3]{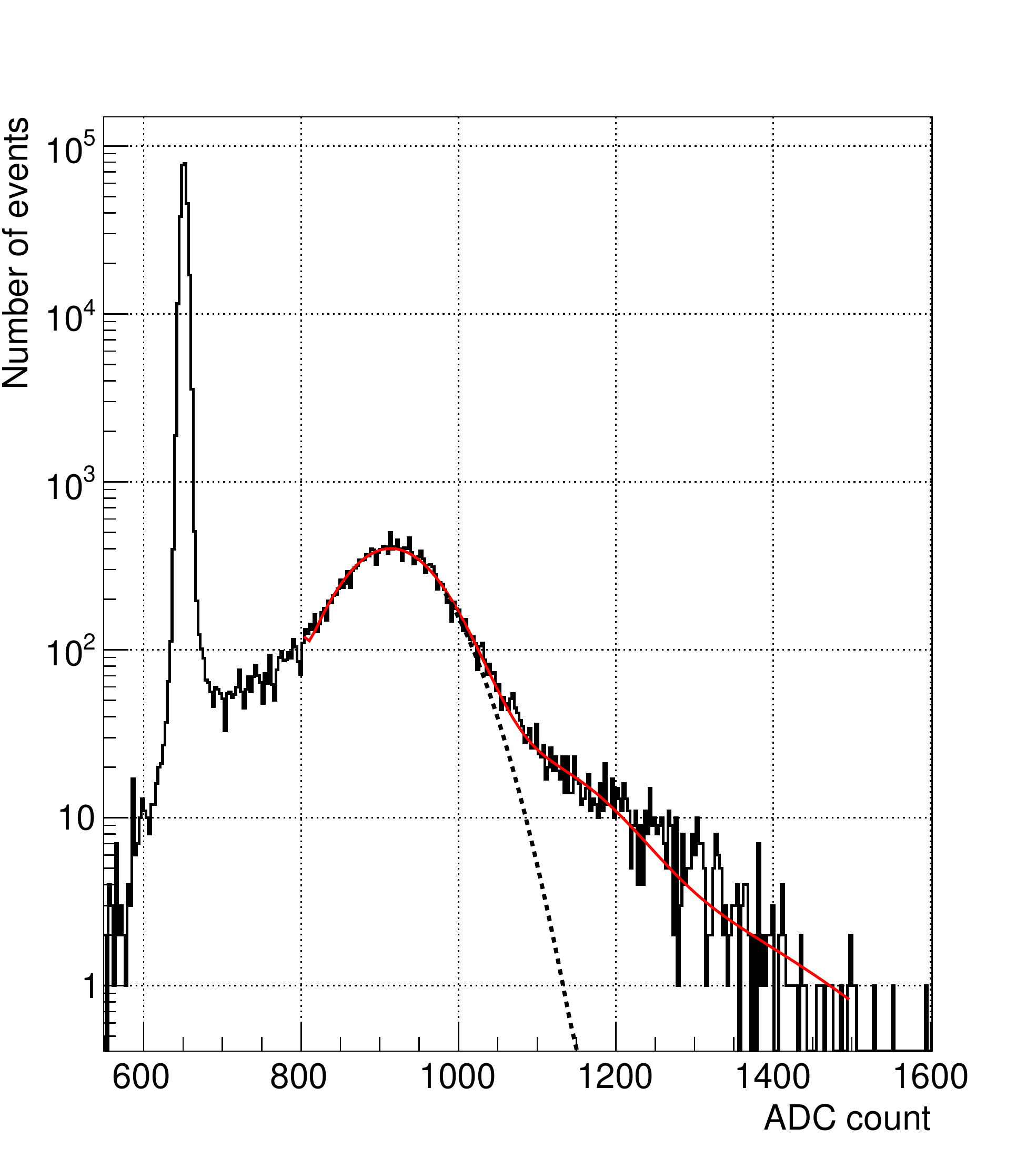}
\caption{With a fit up to several photoelectrons.}
\label{fig:spe_b}
\end{subfigure}

\end{tabular}

\caption{\setlength{\baselineskip}{4mm} ADC distribution of a typical PMT with a fit. (a) Single Gaussian distribution is used for the pedestal peak. (b) For a precise measurement of the PMT gain, the distribution is fitted up to several photoelectrons. The red distribution shows the fit result and the black dashed line indicates the single photoelectron Gaussian distribution from the fit.} 
\label{fig:spe}
\end{figure}

We set the HV to 9 different values between 1200\,V and 1900\,V to measure the gain curve of each PMT. The gain was increased starting from the lowest HV value until the calculated PMT gain was more than 10$^{7}$. For some PMTs, the gain was too small at low values of the HV for the single photoelectron peak to be well-separated from the pedestal. In these cases, the lowest HV values were not used. On average 5 or 6 points were used for each PMT to fit the gain curve. For each HV value, we collected 300,000 events. Figure~\ref{fig:spe} shows the typical ADC count distribution of a PMT with a fit. We separated the fitting range for the pedestal peak and the single photoelectron peak. The fit function models the contributions to the charge spectrum as Gaussian distributions with amplitudes determined by the Poisson distribution up to several photoelectrons to get a precise single photoelectron fit. The fit function is given by
\begin{equation}
\begin{split}
fpedestal(x) = N_{0}~e\frac{-(x-m_{0})^{2}}{2s_{0}^{2}}, \\
fSPE(x) = \sum_{i=1}^{4} N_{i} \frac{\mu^{i} e^{-\mu}}{i!}e\frac{-(x-m_{i})^{2}}{2s_{i}^{2}},   \\
m_{i}  = i \times (m_{1} - m_{0}) + m_{0}, ~
s_{i} = \sqrt i \times s_{1}, 
\end{split}
\end{equation}
where the first line represents a pedestal peak and the second line represents the sum of contributions up to four photoelectrons, $\mu$ is the fraction of the pedestal events, and N$_i$ is the overall normalization. Note that the initial value of the $\mu$ is obtained from the pedestal fit result and treated as a fitting parameter. 

\begin{figure}[h]
\begin{center}
\includegraphics[scale=0.5]{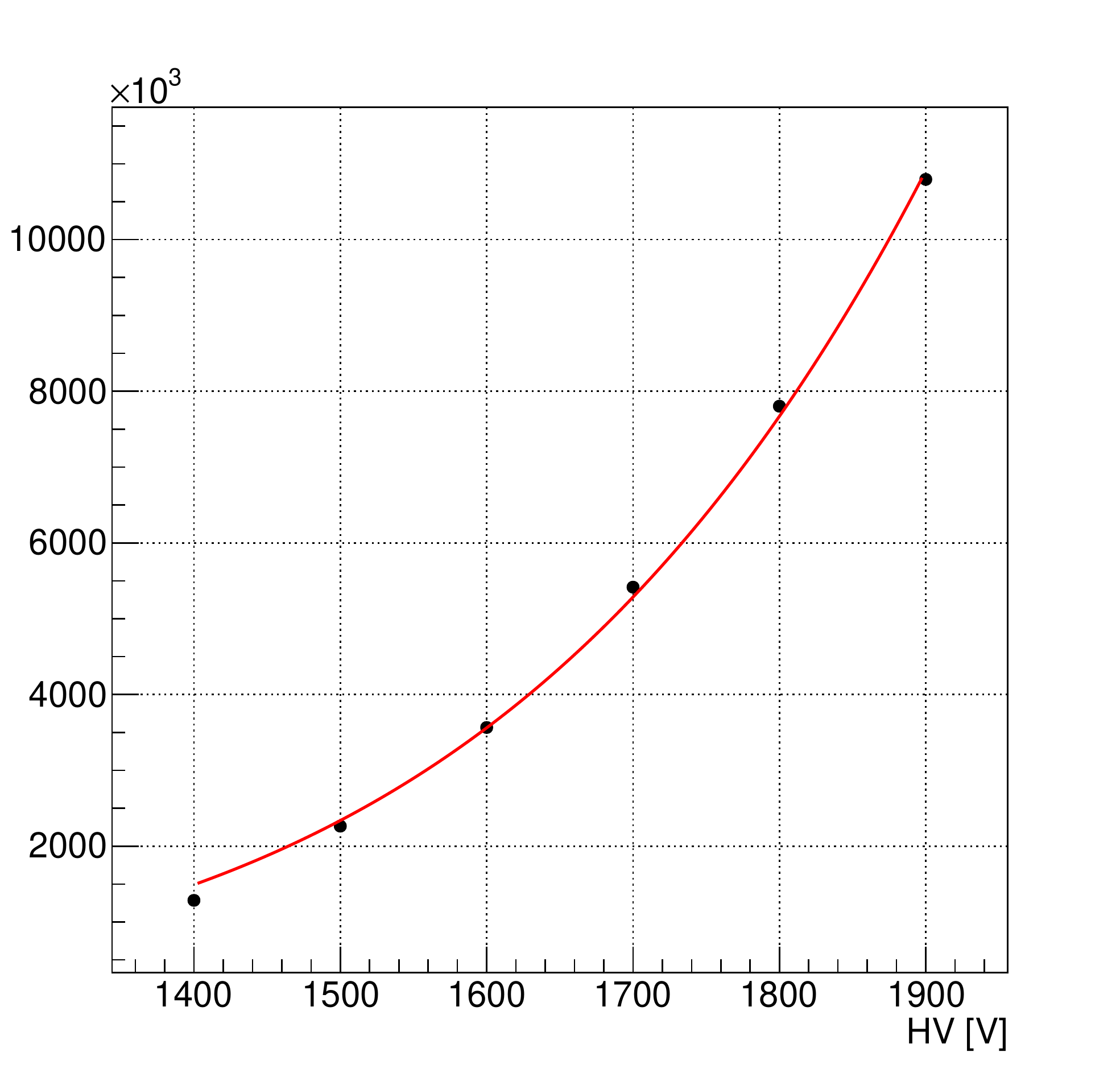}
\end{center}
\caption{\setlength{\baselineskip}{4mm} Measured gain curve of a typical PMT as a function of applied HV (black) and fit (red).}
\label{fig:gain}
\end{figure}

\begin{figure}[h]
\begin{center}
\includegraphics[scale=0.5]{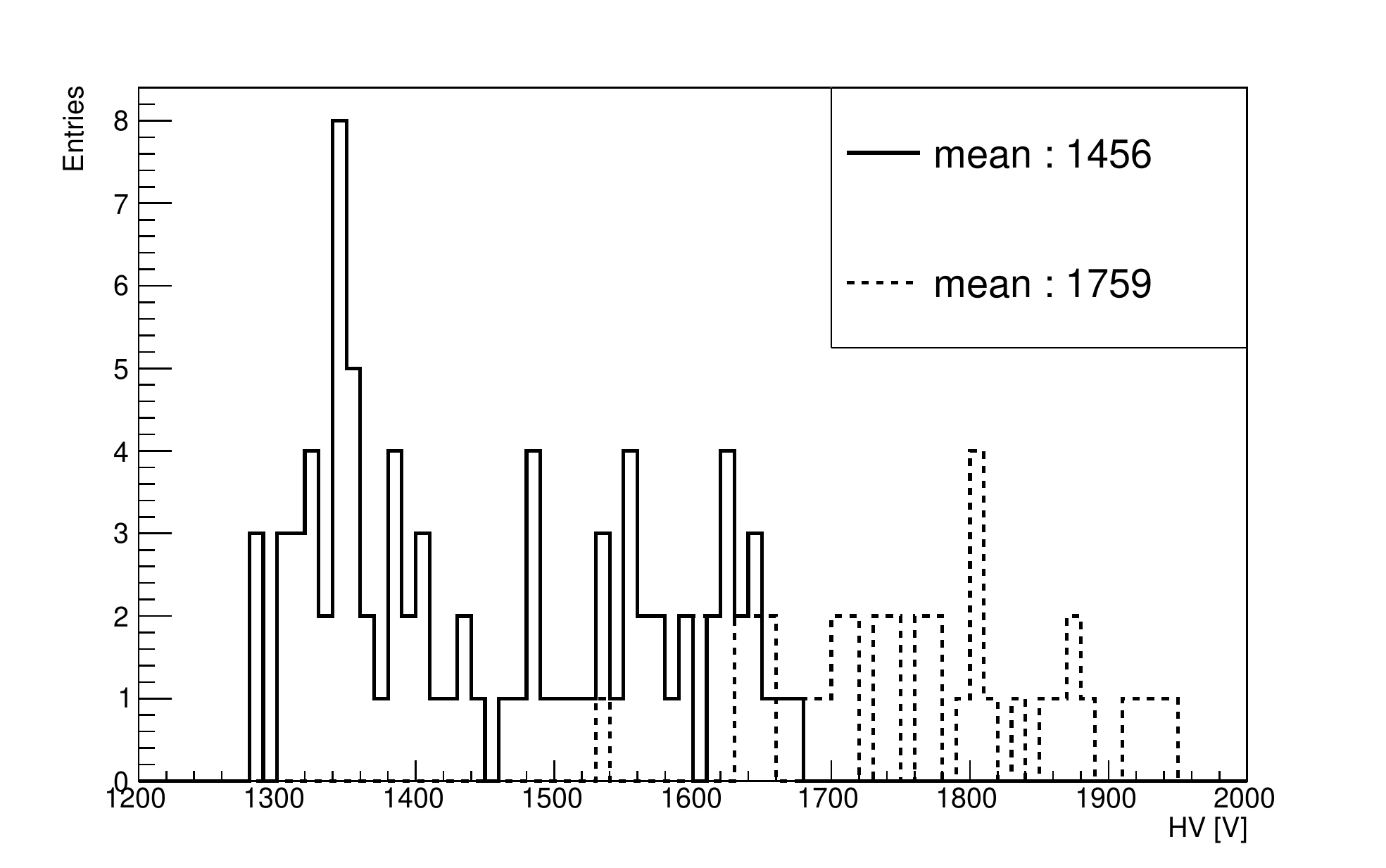}
\end{center}
\caption{\setlength{\baselineskip}{4mm}HV value required to achieve 10$^{7}$ gain for the 123 PMTs. The solid histogram indicates donated PMTs and dashed histogram indicates newly purchased PMTs.}
\label{fig:gain_total}
\end{figure}

The fit result is converted to the PMT gain according to the formula
\begin{equation}
Gain = \frac{2 \times ADC_{count} \times G_{ADC}}{G_{Amp} \times e}, 
\end{equation}
where ADC$_{count}$ is the ADC count difference between the fitted single photoelectron peak and the fitted pedestal peak, G$_{ADC}$ is a conversion factor from ADC counts to charge (0.061 counts/pC in this case), G$_{Amp}$ is the DAQ amplification factor (10 in this case), and $e$ is the electron charge. All PMTs have a 50 Ohm back-terminator, which cuts the analog signal in half, a factor of two is applied in the gain calculation to account for this effect. Figure~\ref{fig:gain} shows the gain curve of one typical PMT with a fit, and Fig.~\ref{fig:gain_total} shows the HV value corresponding to 10$^{7}$ gain for all 123 PMTs which span the range between 1200\,V and 1900\,V. The fit function for the gain curve is well-known from the Hamamatsu~\cite{cite:Hamamatsu_gain} handbook and is given by
\begin{equation}
G = \alpha \times V^{\beta}, 
\end{equation}
where $\alpha$ and $\beta$ are fitting parameters and V is the supplied HV.

\begin{figure}[h]
\begin{center}
\includegraphics[scale=0.5]{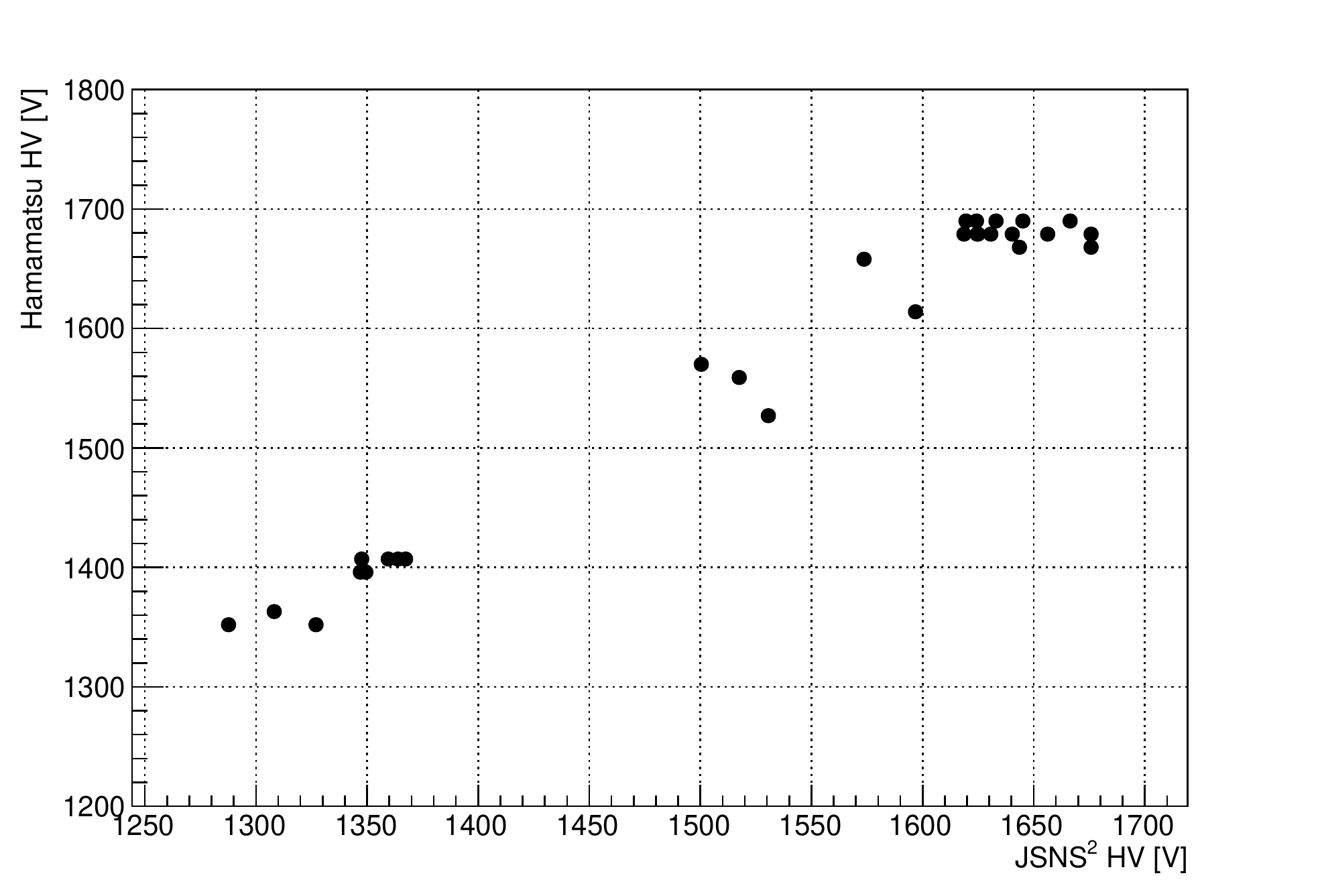}
\end{center}
\caption{\setlength{\baselineskip}{4mm}Comparison between the measured HV value required to achieve 10$^{7}$ gain and the corresponding value from the Hamamatsu data sheets.}
\label{fig:HV_hamamatsu}
\end{figure}

As a cross-check, we compared the measured HV value required to achieve 10$^{7}$ gain for the donated PMTs with the data sheets provided by Hamamatsu. The comparison is shown in Fig.~\ref{fig:HV_hamamatsu} and demonstrates a clear correlation. Note that the Hamamatsu data was only available for a certain subset of the donated PMTs.   

\subsection{Peak-to-Valley ratio at a gain of 10$^{7}$}
\begin{figure}[h]
\begin{center}
\includegraphics[scale=0.5]{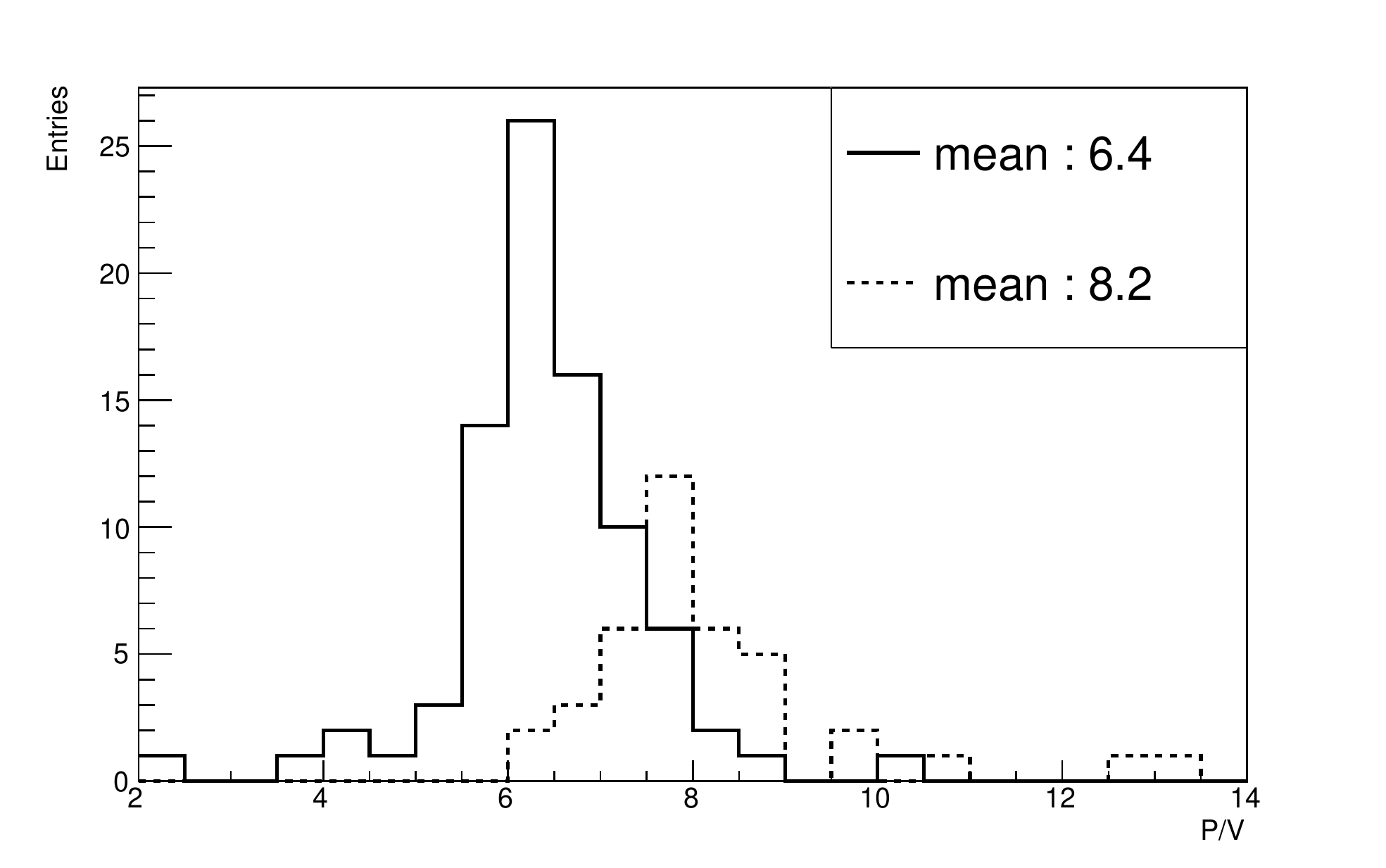}
\end{center}
\caption{\setlength{\baselineskip}{4mm}Peak-to-valley ratio distribution of all 123 PMTs at a gain of 10$^{7}$. The solid histogram corresponds to the donated PMTs and the dashed histogram corresponds to the newly purchased PMTs.}
\label{fig:pv_total}
\end{figure}

The peak-to-valley ratio (P/V) of each PMT at a gain of 10$^{7}$ was calculated with 300,000 laser events. To determine the height of the single photoelectron peak, we fit the PMT charge distribution with the same function used for the PMT gain determination. To estimate the height of the valley, we used the lowest bin value between the pedestal peak and the single photoelectron peak. Figure~\ref{fig:pv_total} shows the P/V distribution of the 123 PMTs.

\subsection{Dark rate measurement at a gain of 10$^{7}$}
\begin{figure}[h]
\begin{center}
\includegraphics[scale=0.5]{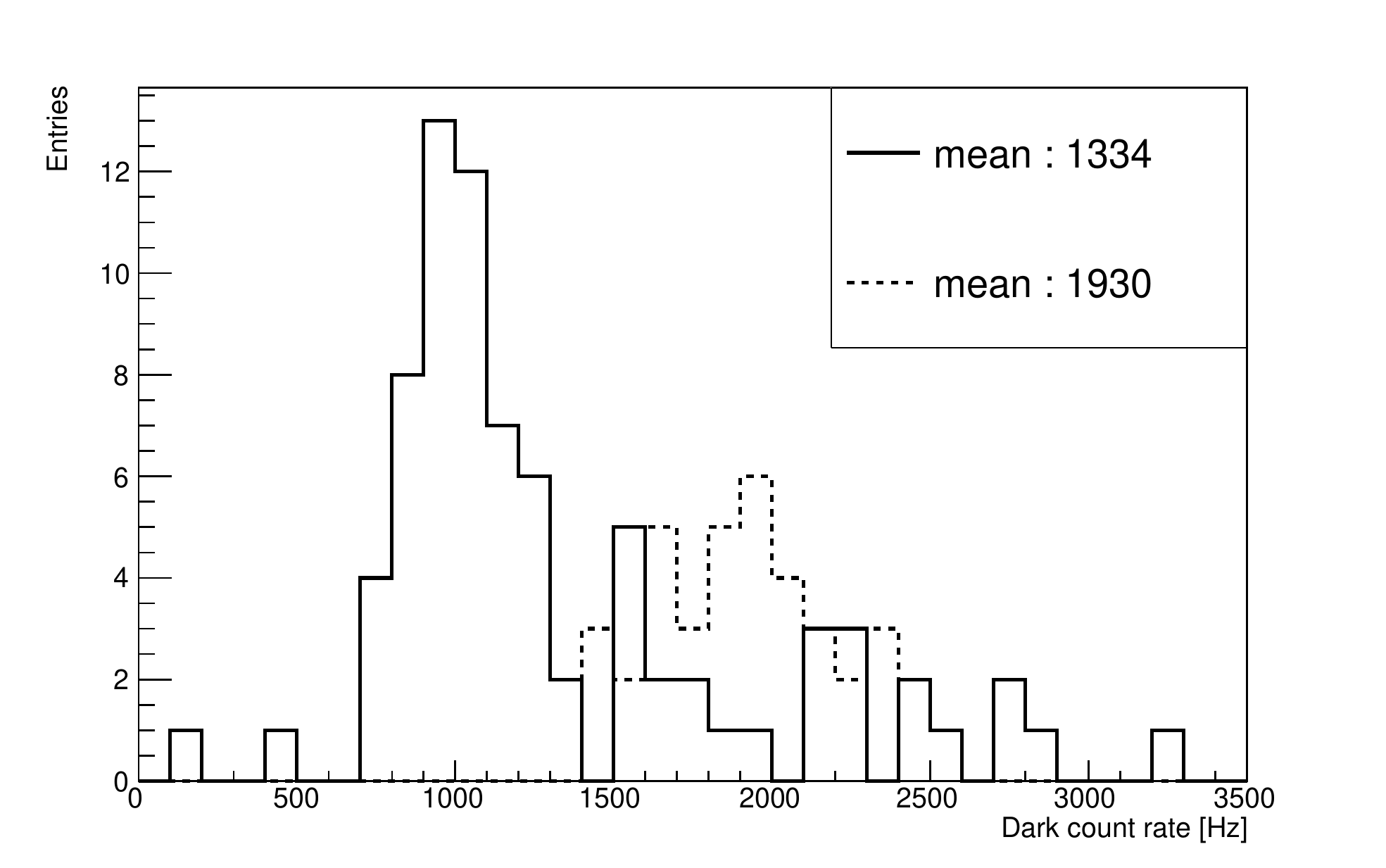}
\end{center}
\caption{\setlength{\baselineskip}{4mm}Dark count rate distribution of all 123 PMTs at a gain of 10$^{7}$. The solid histogram corresponds to the donated PMTs and the dashed histogram corresponds to the newly purchased PMTs.}
\label{fig:dark_total}
\end{figure}

For each PMT, the dark count rate was measured at a gain of 10$^{7}$ using 1 hour of data. The dark count measurements were made approximately 6 hours after the HV was initially applied to the PMTs. We picked the mean value for the dark count rate of each PMT and Fig.~\ref{fig:dark_total} shows the dark count rate distribution of the 123 PMTs. Some of the PMT had dark count rates higher than 4000\,Hz, however, we decided that these PMTs are still suitable for the JSNS$^{2}$ experiment. 

\section{Conclusion}
We performed a PMT pre-calibration campaign for the JSNS$^{2}$ experiment and found that 123 PMTs performed satisfactorily. For the PMTs where the Hamamatsu data sheets were available, we compared the measured HV value required to achieve 10$^{7}$ gain with the data sheets and found a clear correlation. For comparison, table 3 of Ref~\cite{cite:DC_PMT} shows the PMT pre-calibration results of the Double Chooz experiment. In the table, the HV at a gain of 10$^{7}$ is between 1210\,V and 1610\,V, the maximum value of the P/V is 5.5, and the maximum value of the dark count rate is 9000\,Hz. The JSNS$^{2}$ PMT pre-calibration results show similar or better performance, noting that some differences may be due to the different measurement conditions and equipment such as a 14-bit ADC and a laser module as a light source. We concluded that the performance of the PMTs, including the PMTs donated from other experiments, is good enough for the JSNS$^{2}$ experiment. In light of this, we installed 120 PMTs in the detector and saved 3 PMTs as backups.  

\section*{Acknowledgment}
We thank the J-PARC staff for their support. We acknowledge the support of the Ministry of Education, Culture, Sports, Science and Technology (MEXT) and the JSPS grants-in-aid (Grant Number 16H06344, 16H03967), Japan. This work is also supported by the National Research Foundation of Korea (NRF) Grant No. 2016R1A5A1004684, 2017K1A3A7A09015973, 2017K1A3A7A09016429, 2019R1A2C3004955, 2016R1D1A3B02010606, 2017R1A2B4011200 and 2018R1D1A1B07050425 funded by the Korea Ministry of Science and ICT. Our work have also been supported by a fund from the BK21 of the NRF. The University of Michigan gratefully acknowledges the support of the Heising-Simons Foundataion.This work conducted at Brookhaven National  Laboratory was supported by the U.S. Department of Energy under Contract DE-AC02-98CH10886.The work of the University of Sussex is supported by the Royal Society grant no.IES\textbackslash R3\textbackslash 170385.


\begin{thebibliography}{99}

\bibitem{cite:JSNS2_proposal}
JSNS$^{2}$ Proposal, M. Harada et al., https://arxiv.org/abs/1310.1437

\bibitem{cite:JSNS2_TDR}
JSNS$^{2}$ TDR, S. Ajimura et al., https://arxiv.org/abs/1705.08629

\bibitem{cite:JSNS2_Veto_detector}
Y. Hino et al., JINST 14 (2019) no.09 T09001

\bibitem{cite:DayaBay_GdLS}
M. Yeh et al., Nuclear Instruments and Methods in Physics Research A 584 (2008) 238 - 243

\bibitem{cite:R7081}
https://www.hamamatsu.com/resources/pdf/etd/LARGE\_AREA\_PMT\_TPMH1376E.pdf

\bibitem{cite:Hamamatsu_gain}
https://www.hamamatsu.com/resources/pdf/etd/PMT\_handbook\_v3aE.pdf

\bibitem{cite:DC_PMT}
T. Matsubara et al., Nuclear Instruments and Methods in Physics Research A 661 (2012) 16 - 25

\end{thebibliography}
\end{document}